\documentclass{article}

\usepackage{arxiv}

\usepackage[utf8]{inputenc} 
\usepackage[T1]{fontenc}    
\usepackage{hyperref}       
\usepackage{url}            
\usepackage{booktabs}       
\usepackage{amsfonts}       
\usepackage{nicefrac}       
\usepackage{microtype}      
\usepackage{graphicx}
\usepackage{doi}
\usepackage{multirow} 
\usepackage{listings}

\title{Efficiency of nonparametric superiority tests based on restricted mean survival time versus the log-rank test under proportional hazards }

\author{ \href{https://orcid.org/0000-0002-3959-7057}{\includegraphics[scale=0.06]{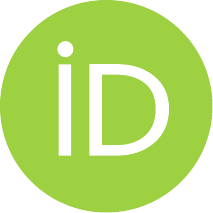}\hspace{1mm}Dominic Magirr}\\
	Advanced Quantitative Sciences\\
	Novartis Pharma AG\\
    Basel, Switzerland \\
	\texttt{dominic.magirr@novartis.com} \\
	\And
	\href{https://orcid.org/0000-0003-1804-2463}{\includegraphics[scale=0.06]{orcid.pdf}\hspace{1mm}Craig Wang} \\
	Advanced Quantitative Sciences\\
	Novartis Pharma AG\\
    Basel, Switzerland \\
	\texttt{craig.wang@novartis.com} \\
 \And
	\href{https://orcid.org/0000-0001-8129-6007}{\includegraphics[scale=0.06]{orcid.pdf}\hspace{1mm}Xinlei Deng} \\
	Advanced Quantitative Sciences\\
	Novartis Pharma AG\\
    London, UK \\
	\texttt{xinlei.deng@novartis.com} \\
     \And
	\href{https://orcid.org/0000-0001-5850-3610}{\includegraphics[scale=0.06]{orcid.pdf}\hspace{1mm}Tim P.\ Morris} \\
	MRC Clinical Trials Unit at UCL\\
    London, UK \\
	\texttt{tim.morris@ucl.ac.uk} \\
 \And
	\href{https://orcid.org/0000-0002-5618-0667}{\includegraphics[scale=0.06]{orcid.pdf}\hspace{1mm}Mark Baillie} \\
	Advanced Quantitative Sciences\\
	Novartis Pharma AG\\
    Basel, Switzerland \\
	\texttt{mark.baillie@novartis.com} \\
}

\renewcommand{\shorttitle}{\textit{arXiv} Template}

\hypersetup{
pdftitle={Efficiency of tests based on RMST vs. log-rank},
pdfsubject={Efficiency of tests based on RMST vs. log-rank},
pdfauthor={Dominic Magirr, Craig Wang, Xinlei Deng, Tim P. Morris, Mark Baillie},
pdfkeywords={Hazard ratio, Restricted mean survival, Proportional hazards},
}

\begin{document}
\maketitle

\begin{abstract}
\textbf{Background} For randomized clinical trials with time-to-event endpoints, proportional hazard models are typically used to estimate treatment effects and log-rank tests are commonly used for hypothesis testing. The summary measure of the primary estimand is frequently a hazard ratio. However, there is growing support for replacing this approach with a model-free estimand and assumption-lean analysis method—a trend already observed for continuous and binary endpoints. One alternative is to base the analysis on the difference in restricted mean survival time (RMST) at a specific restriction time, a single-number summary measure that can be defined without any restrictive assumptions on the outcome model. In a simple setting without covariates, an assumption-lean analysis can be achieved using nonparametric methods such as Kaplan--Meier estimation. The main advantage of moving to a model-free summary measure and assumption-lean analysis is that the validity and interpretation of conclusions do not depend on the proportional hazards (PH) assumption. The potential disadvantage is that the nonparametric analysis may lose efficiency under PH. There is disagreement in recent literature on this issue, with some studies indicating similar efficiency between the two approaches, while others highlight significant advantages for PH models.

\textbf{Methods} Existing asymptotic results and a simulation study are used to compare the efficiency of a log-rank test against a nonparametric analysis of the difference in RMST in a superiority trial under PH. Previous studies have separately examined the effect of event rates and the censoring distribution on relative efficiency. This investigation clarifies conflicting results from earlier research by exploring the joint effect of event rate and censoring distribution together. Several illustrative examples are provided.

\textbf{Results} In scenarios with high event rates and/or substantial censoring across a large proportion of the study window, and when both methods make use of the same amount of data, relative efficiency is close to unity. However, in cases with low event rates but when censoring is concentrated at the end of the study window, the proportional hazards analysis has a considerable efficiency advantage. 

\textbf{Conclusions} There are realistic combinations of event rates and censoring distributions where, if the proportional hazards assumption holds, the proportional hazards analysis is more efficient than a nonparametric RMST-based test. Additionally, efficiency of the proportional hazards analysis can benefit from data collected beyond the restriction time. A key take-away message is that the implications of requiring assumption-lean analysis methods should be carefully considered during the trial design phase.
\end{abstract}


\section{Introduction}

A movement towards model-free estimands and assumption-lean analysis methods in the primary analysis of RCTs (randomized clinical trials) has been gathering pace in recent years \cite{van2015statistics, mutze2024principles, chen2023beyond, vansteelandt2024assumption, bartlett2020hazards, mccaw2021choosing}. Already, for continuous and binary outcomes, it could be considered standard practice, exemplified by the FDA guidance document on covariate adjustment \cite{fda2023adjusting}. However, for trials with time-to-event outcomes, it is still common to see analysis methods with strong modelling assumptions, in particular, assumptions regarding proportional hazards. This is despite an increasing number of publications spelling out the limitations of such an approach \cite{hernan2010hazards, stensrud2020test, martinussen2020subtleties, bartlett2020hazards, fay2024causal, heitjan2024comment, fay2024reply}.

The motivation for moving to a model-free estimand and assumption-lean analysis for time-to-event endpoints is clear. The result of the primary analysis does not rely on the possibly incorrect proportional hazards (PH) assumption (defined in section \ref{secMethods}). 

One potential disadvantage of moving to a model-free estimand and assumption-lean analysis method is a loss in efficiency compared to a proportional hazards analysis. However, previous research on Kaplan--Meier based estimation of differences in restricted mean survival time (RMST) suggested that this particular model-free, assumption-lean approach incurs minimal efficiency loss compared to a PH analysis, even under PH \cite{tian2018efficiency, eaton2020designing, horiguchi2020empirical, royston2013restricted}. An outlier in this line of research is the analysis of Freidlin et al.\ \cite{freidlin2021restricted}, who found substantial efficiency advantages for the PH analysis, prompting questions about what was different in their investigation. In this paper, we aim to give a thorough explanation for these prior conflicting results.
    
Relative efficiency of statistical tests for time-to-event outcomes is a broad and well studied topic \cite{meier2004price}. Our scope will be limited in the following ways. Firstly, we only look at superiority studies. Special considerations apply to non-inferiority studies \cite{quartagno2021restricted, freidlin2021reply}. We also do not consider covariate-adjusted analysis. Adjustment for baseline prognostic covariates is known to improve efficiency of both nonparametric-RMST and Cox analysis \cite{diaz2019improved, wei2024conditional, ye2024covariate, hernandez2006randomized, kahan2014risks}. Understanding  relative efficiency in the simpler situation without covariates is a logical first step before attempting to characterize relative efficiency of covariate-adjusted methods. Furthermore, we limit our investigation to scenarios where the PH assumption is correct, as this is where there is most disagreement in the literature. For non-PH scenarios, there is broad consensus that nonparametric RMST-based tests tend to be more efficient than a PH analysis in “early effect” scenarios (i.e.\ where the treatment effect is observed shortly after treatment initiation), whereas the PH analysis tends to be more efficient in “late effect” scenarios \cite{tian2018efficiency, gregson2019nonproportional, jimenez2024visualizing, klinglmuller2023neutral,eaton2020designing}. 

Our focus is on relative statistical efficiency, rather than the relative ease of interpretation of different summary measures. The comparison that we are making, between nonparametric estimation of differences in RMST and semi-parametric estimation of a hazard ratio (HR) using a Cox model, is challenging in the sense that we are altering two things at the same time, both the estimand and the estimation method. We could, alternatively, have chosen to fix the estimand to be the difference in RMST, and investigated the relative efficiency of non-parametric versus semi-parametric estimation \textit{via} a Cox model. Such a comparison would be cleaner. One would also expect the model-based RMST analysis to inherit both the efficiency and the vulnerability to model misspecification of the underlying Cox model \cite{quartagno2023comparison}, meaning that similar conclusions would be drawn for the nonparametric RMST \textit{vs}.\ semiparametric RMST comparison as for the nonparametric RMST \textit{vs}.\ semiparametric HR comparison. Nevertheless, we focus on the  nonparametric RMST \textit{vs}.\ semiparametric HR comparison as it is more obviously relevant to the current debate, which is about whether to move away from the status quo of log-rank tests and estimation of hazard ratios towards model-free estimands and assumption-lean analysis methods.

\section{Methods}\label{secMethods}

\subsection{Model-free estimands, assumption-lean analysis}

 For RCTs with a time-to-event endpoint, we are often willing to assume that the vectors $(T_i(1), T_i(0))$ of potential survival times on the two treatments, for patients $i=1,\ldots, N$, are independent identically distributed random variables from some distribution $f$. By “model-free estimand", we mean that the estimand is not predicated on any particular statistical modeling assumptions beyond this basic one, and possibly some regularity conditions needed for existence of moments, etc. In other words, the estimand is some functional of $f$. An example could be the difference in means of the potential outcomes, $E_f(T(1) - T(0))$, or the difference in medians, etc. The \textit{hazard ratio function},
\begin{equation}
    \mathrm{HR}(t) = \frac{f_1(t)}{S_1(t)}\frac{S_0(t)}{f_0(t)},
\end{equation}
is a functional of $f$, with $f_j(t)$ and $S_j(t)$ the marginal density and survival distributions of the potential outcomes $T(j)$ for $j = 0,1$. By our definition, it is a model-free estimand. It is not, however, a single-number summary measure. Arguably, a more prevalent estimand is a single-number \textit{hazard ratio}  predicated on the assumption of proportional hazards, $\mathrm{HR}(t) =  \mathrm{HR}$ for all $t$. This would be considered a “model-based estimand".

 Owing to randomization, we assume that treatment assignment $A_i \in \{0,1\}$ is independent of the potential outcomes $(T_i(1), T_i(0))$. In addition, we are often willing to assume a censoring mechanism that is independent of outcome, possibly conditional on treatment arm (although informative censoring is often an important issue in practice). By “assumption-lean analysis", we mean an analysis method whereby valid inference (e.g., confidence intervals that have correct coverage asymptotically) requires these rather minimal assumptions, together with the basic ones above, but does not require further particular statistical assumptions on $f$, such as proportional hazards, or any kind of parametric form. 

 Regarding the use of proportional hazards models to estimate hazard ratios, one could view this approach as targeting a model-free estimand, $\mathrm{HR}(t)$. Having defined the estimand in a model-free way, one proceeds to estimate $\mathrm{HR}(t)$ \textit{via} an assumption-heavy analysis. Alternatively, if the estimand is defined as a single-number $\mathrm{HR}$ predicated on the PH assumption, one might call this a model-based estimand. 
 It is not particularly important which of these subtly different perspectives is taken. Importantly, it is the combination of a model-free estimand and an assumption-lean analysis method that is advocated as a robust approach to the primary analysis of RCTs.

\subsection{RMST}
Using similar notation to Tian \textit{et al.}~\cite{tian2018efficiency},  the RMST is defined as $E(\min\{ T, \tau \})$, where $T$ denotes the event time and $\tau$ is the restriction time. In a two arm trial, the between-group difference in RMST, denoted $D$, is equal to the area between two survival curves over $[0, \tau]$,
\begin{equation} \label{eqRMST}
    D = \int_0^{\tau}  \left\lbrace S_1(t) - S_0(t) \right\rbrace dt, 
\end{equation}
where $S_j$ denotes the survival distribution on treatment $j=0,1$. This is a model-free estimand. An assumption-lean analysis can be performed based on the estimator, $\hat{D}$, which is formed by plugging in Kaplan--Meier estimates $\hat{S}_j$ into (\ref{eqRMST}). A consistent estimator for $\mathrm{var}(\hat{D})$ is available (eq. 1 of \cite{tian2018efficiency}), \cite{pepe1989weighted, uno2022package}.

The statistic $\hat{D}$ could be used to test a null hypothesis $D=0$. But it could also be used to test a null hypothesis that the survival distributions on the two arms are identical. If the survival distributions are identical then $D=0$, but not necessarily vice versa. To facilitate comparison with the log-rank test, we will focus on the null hypothesis of identical survival functions. To further facilitate an asymptotic comparison with the log-rank test (as will be made clear in the following sections), we can make use of the result from Tian et al. \cite{tian2018efficiency} that, for alternative hypotheses close to the null (see \cite{tian2018efficiency} for a precise definition), the RMST-based test is asymptotically equivalent to a test based on 
\begin{equation}\label{eqAsD}
    \int_0^{\tau} \frac{\int_t^{\tau} S_0(v) dv}{\int_0^\tau S_0(v)dv}d\left\lbrace \widehat{\Lambda}_1(t) - \widehat{\Lambda}_0(t) \right\rbrace,
\end{equation}
where $\widehat{\Lambda}_j$ denotes the Nelson--Aalen estimator of the cumulative hazard function on treatment $j=0,1$. 
\subsection{PH analysis}
For the PH analysis, let $\hat{\theta}$ be the estimator of the log hazard ratio, $\theta$, which is assumed to be constant. The estimator $\hat{\theta}$ is found by maximizing the partial likelihood function of a Cox model with treatment term only. Again, following Tian et al. \cite{tian2018efficiency}, for alternative hypotheses close to the null, a test based on $\hat{\theta}$ such as the log-rank test is asymptotically equivalent to a test based on 
\begin{equation}\label{eqAsTheta}
    \int_0^{t_H} S_C(t)S_0(t)d\left\lbrace \widehat{\Lambda}_1(t) - \widehat{\Lambda}_0(t) \right\rbrace,
\end{equation}
where $t_H$ denotes the end of follow-up for the log-rank test, and $S_C$ denotes the censoring distribution. For the purpose of our comparison, we shall assume the censoring distribution is common across both arms, although this assumption is not typically needed for the log-rank test.

\subsection{Relative efficiency}
As can be seen from (\ref{eqAsD}) and (\ref{eqAsTheta}), for situations where $\tau \approx t_H$, the asymptotic relative efficiency of the tests (RMST and PH analysis) only depends on the weight functions
\begin{equation}\label{eqWD}
    w_D(t)=\frac{\int_t^\tau S_0(v) dv}{\int_0^\tau S_0(v)dv}
\end{equation}
and 
\begin{equation}\label{eqWtheta}
    w_\theta(t)=S_C(t)S_0(t) 
\end{equation}
for $t \in [0, \tau]$. Given assumptions about the control survival and censoring distributions, one need only plot these two weight functions in order to understand the relative behavior of the test statistics. Since the log-rank test is the optimal rank-based test under PH, one would expect the weights (\ref{eqWtheta}) to be close to optimal for these scenarios. If (\ref{eqWD}) is giving larger weights to earlier timepoints, relative to (\ref{eqWtheta}), then we would expect this to be less efficient under PH, and also for scenarios where the hazard ratio is improving over time, but more efficient for scenarios when the hazard ratio is diminishing over time. 

Examination of  (\ref{eqWD}) and (\ref{eqWtheta}) will form the first part of our investigations, based on the same scenarios that will be used in our simulation study, as described below.

\subsection{Simulation study}

While investigation of the weight functions (\ref{eqWD}) and (\ref{eqWtheta}) is helpful to gain a qualitative sense of the relative behavior of the two test statistics, it gives no sense of the magnitude of power differences. For that, we use a simulation study.

\begin{table*}[h]
\centering
\begin{tabular}{| c |c c |  c c  c c  c | c c|}
\hline
Scenario & Event rate       & $S_1(3)$            & HR                  & $n_j$ &  $\tau$ & $t_H$ & $t^+_{H}$ & Recruitment & $t_R$  \\
\hline 
1&\multirow{4}{*}{Low}      &\multirow{4}{*}{0.9}  & \multirow{4}{*}{0.67}  & \multirow{4}{*}{1,000} &  \multirow{4}{*}{3} &  \multirow{4}{*}{3}  &  \multirow{4}{*}{3.5}& Instant     & 0      \\
2&                          &                      &                        &   &   & & & Fast        & 0.5  \\
3&                          &                       &                        &   &   & & & Moderate    & 1.5 \\
4&                          &                       &                        &   &   & && Slow        & 2.5  \\  
\hline
5&\multirow{4}{*}{Moderate} &\multirow{4}{*}{0.6}   & \multirow{4}{*}{0.67}  & \multirow{4}{*}{250}  & \multirow{4}{*}{3}    & \multirow{4}{*}{3}  &  \multirow{4}{*}{3.5} & Instant     & 0    \\
6&                          &                       &                        &   &   &  && Fast        & 0.5 \\
7&                          &                      &                        &   &   &  & & Moderate    & 1.5 \\
8&                          &                     &                        &   &   &  &  & Slow        & 2.5 \\
\hline
9&\multirow{4}{*}{High}     &\multirow{4}{*}{0.2}   & \multirow{4}{*}{0.67}  & \multirow{4}{*}{150} & \multirow{4}{*}{3}    & \multirow{4}{*}{3}   &  \multirow{4}{*}{3.5}  &   Instant   & 0   \\
10&                         &                     &                        &   &     &  &  &   Fast      & 0.5   \\
11&                         &                     &                        &   &     &  &  &   Moderate  & 1.5   \\
12&                         &                     &                        &   &     &  &   &   Slow      & 2.5  \\
\hline
\end{tabular}
\caption{\label{tabScenarios} Simulation scenarios; $t_R$ is the duration of the recruitment period, $n_j$ the sample size per arm, $\tau$ the restriction time of the RMST analysis, $t_H$ the end of follow-up for the log-rank test when both methods make use of the same amount of data, and $t_H+$ the end of follow-up for the log-rank test when it makes use of additional data after $\tau$.}
\end{table*}

The aim of the simulation study is to compare the power of a non-parametric analysis of difference in RMST with that of a log-rank test for testing a null hypothesis of equal survival distributions. Our scenarios focus on proportional hazards, but we vary both the event rates and the recruitment rates across realistic ranges. We seek to characterize those scenarios where relative efficiency is close to one, versus those scenarios where the power differs, according to the combination of event rate and recruitment rate. In addition, we seek to quantify the magnitude of power differences.

\begin{figure*}
\centering
\includegraphics[width=\linewidth]{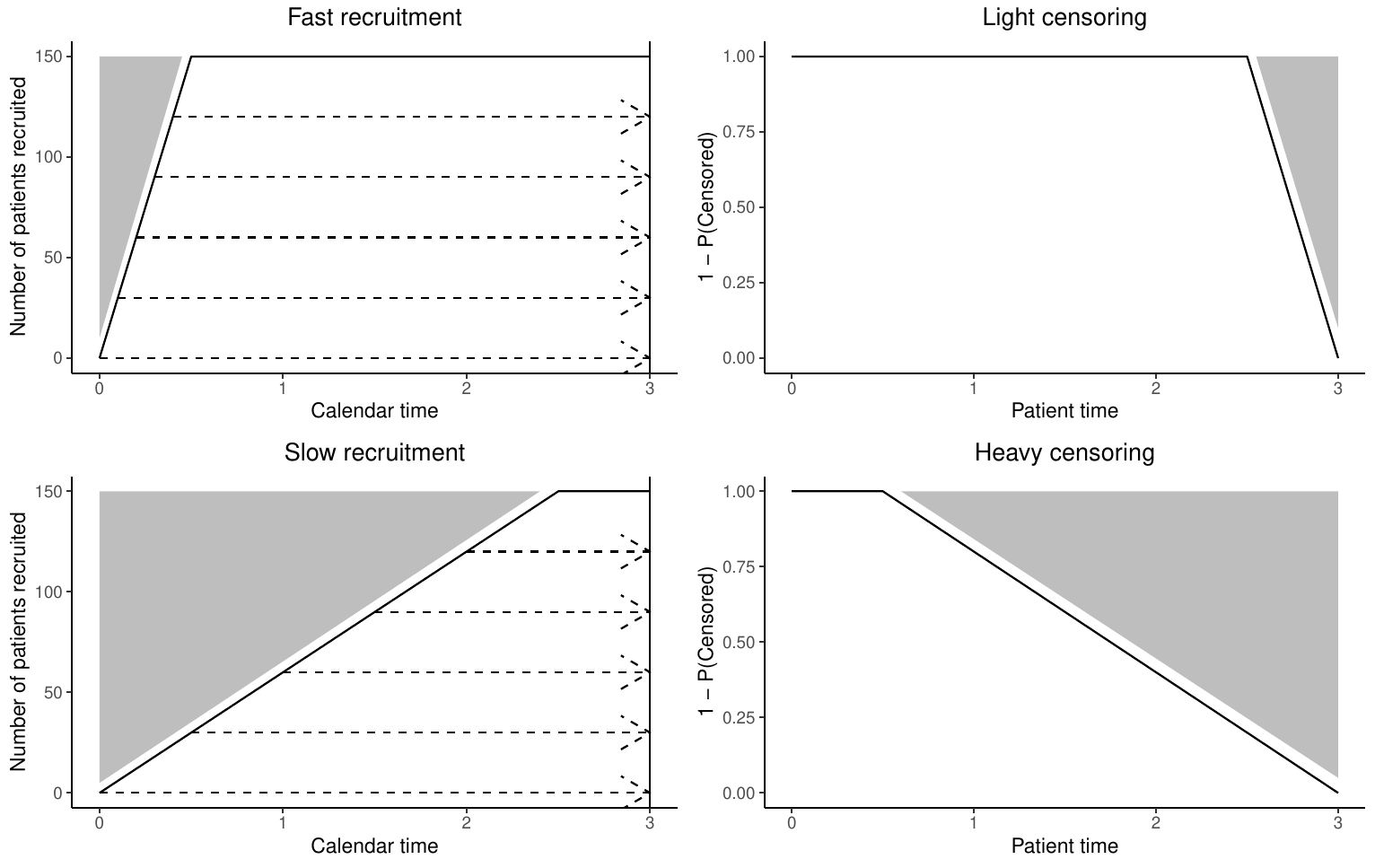}
\caption{\label{figCensoring} Illustration of how a fast recruitment rate (upper left) leads to a “light” censoring distribution (upper right), and a slow recruitment rate (lower left) leads to a “heavy” censoring distribution (lower right), when the only form of censoring is administrative censoring at the end of the study follow-up.}
\end{figure*}

The scenarios we consider (Table~\ref{tabScenarios}) are an extension of those in Freidlin et al. \cite{freidlin2021restricted}. In the first set of scenarios considered by Freidlin et al., all patients are followed up for 3 years, which also corresponds to the restriction time of the RMST-based test. We refer to this scenario as “Instant” recruitment because this is the censoring pattern that would be produced in an event-driven trial (where the data cut-off occurs after a pre-specified number of events) had the recruitment been instantaneous and had administrative censoring (due to end of study) been the only source of censoring. Interestingly, Freidlin et al.\cite{freidlin2021restricted} chose this censoring pattern as they anticipated it gave the most advantage to the RMST-based test. We supplement this censoring pattern with three other assumptions about the recruitment duration (denoted $t_R$) corresponding to “Fast", “Moderate” and “Slow” recruitment, each with a constant recruitment rate, where we fix the total trial duration at $t_H =3$ years and administrative censoring is the only source of censoring. Figure \ref{figCensoring} illustrates how a fast recruitment rate leads to a light censoring distribution while a slow recruitment rate leads to a heavy censoring distribution. In all cases, the probability of remaining uncensored at 3 years is zero. So, in our context, when we talk about “light” censoring, we do not mean that the overall probability of being censored at the end of the study is low. Rather, we mean that the probability of being censored is very low for the majority of the study window, with censoring concentrated at the end of the study window. Conversely, when we talk about “heavy” censoring, we mean that there is a non-trivial probability of being censored across a large proportion of the study window. The “Moderate” recruitment scenario gives a similar censoring distribution to the one used in Tian et al. \cite{tian2018efficiency}. Like Freidlin et al., we consider exponentially distributed survival with three different event rates: a “Low” event rate scenario where the survival probability in the experimental arm after 3 years is $0.9$, as well as a “Moderate” event rate ($S_1(3) = 0.6$) and a “High” event rate ($S_1(3) = 0.2$). The sample size per arm ($n_j$) is tied to the event rate in order to produce power somewhere between $0.8$ and $0.9$ for the log-rank test under instant recruitment. For all combinations of event rate and recruitment rate, we consider a hazard ratio of $0.67$. If, for a particular simulated data set, one of the study arms has no patients left in the risk set at 3 years, then we cannot reliably estimate the difference in RMST at 3 years without introducing further assumptions. In this case, we change $\tau$ to be the minimum of the maximum follow-up time on each arm, so that nonparametric estimation can still be used (though this also depends on certain assumptions about the censoring distribution \cite{tian2020empirical}). For each scenario where we start with a target $\tau = t_H = 3$, we will report the average value, $\bar{\tau}$, based on this tweak, which we expect to still be close to $3$. The log-rank test does not use any information beyond $\tau$, even with this tweak. We note that changing the estimand in response to numerical difficulties in estimation is an important practical issue, hence the need for the additional scenarios that we describe next.

Scenarios where $\tau = t_H$ are in a sense  best-case scenarios in terms of the efficiency of the nonparametric RMST-based test under PH. When $\tau = t_H$, both methods make use of the same amount of data. While this might appear to be a natural and fair comparison, it is not necessarily a fair reflection of “real-world” efficiency. In the context of the primary analysis of an RCT, $\tau$ may need to be pre-specified to a so-called clinically meaningful timepoint, and it may not be possible to tweak this value of $\tau$ to the last available observation time. Depending on how the actual recruitment, censoring and event rates correspond to those used in the planning phase, there may well be additional data available after $\tau$, which could be used in a log-rank test but not in a non-parametric analysis of RMST with the pre-fixed $\tau$.  Freidlin et al. attempted to capture this issue with a second set of scenarios where the total trial length is fixed at 6 years, while the recruitment duration and restriction time for the RMST-based test are fixed at 3 years. Arguably, however, if discrepancies between planned and realized event/recruitment rates are very large, then this would likely precipitate a protocol amendment, regardless of the analysis method. We focus, therefore, on a second set of scenarios where $t_H$ is larger than $\tau$ by a modest amount. Specifically, we change the follow-up time for the log-rank test from $t_H=3$ to $t_H^+=3.5$. The aim is to reflect the realistic scenario that a small amount of follow-up data is available after $\tau$ but not enough to precipitate a protocol amendment to increase $\tau$. This should give a better reflection of real-world efficiency, with the acknowledgment that this is difficult to quantify in a neutral way.

As well as estimating the power for the RMST-based and log-rank tests (denoted by $\Pi_D$ and $\Pi_\theta$, respectively)  for each scenario, we will also turn this into an approximate measure of relative efficiency ($\mathrm{R.E.}$), as follows,
\begin{equation}
\mathrm{R.E.} = \left(\frac{\Phi^{-1}(0.975) + \Phi^{-1}(\Pi_D)}{\Phi^{-1}(0.975) + \Phi^{-1}(\Pi_\theta)}\right)^2.
\end{equation}
Roughly speaking, the $\mathrm{R.E.}$ tells us the relative sample size needed to achieve the same power using the two tests. When $\mathrm{R.E.} < 1$, this tells us that the PH analysis would require a smaller size.

\textit{Implementation}

For each scenario, we generate 10,000 repetitions of the data and analysis. For the RMST analysis, we use the R package survRM2 \cite{therneau2013r}. For the log-rank test, we use the survival package \cite{uno2015vignette}. Code to reproduce the results is available at \url{https://github.com/dominicmagirr/efficiency_of_rmst_paper}.
All simulations are implemented under R version 4.3.1.

\subsection{Case studies}

To supplement our asymptotic and simulation evidence, with the aim of reinforcing our understanding about the relative behavior of the RMST- and PH-based tests, we present four case studies, which we shall briefly describe here.

\begin{figure*}
\centering
\includegraphics[width=\linewidth]{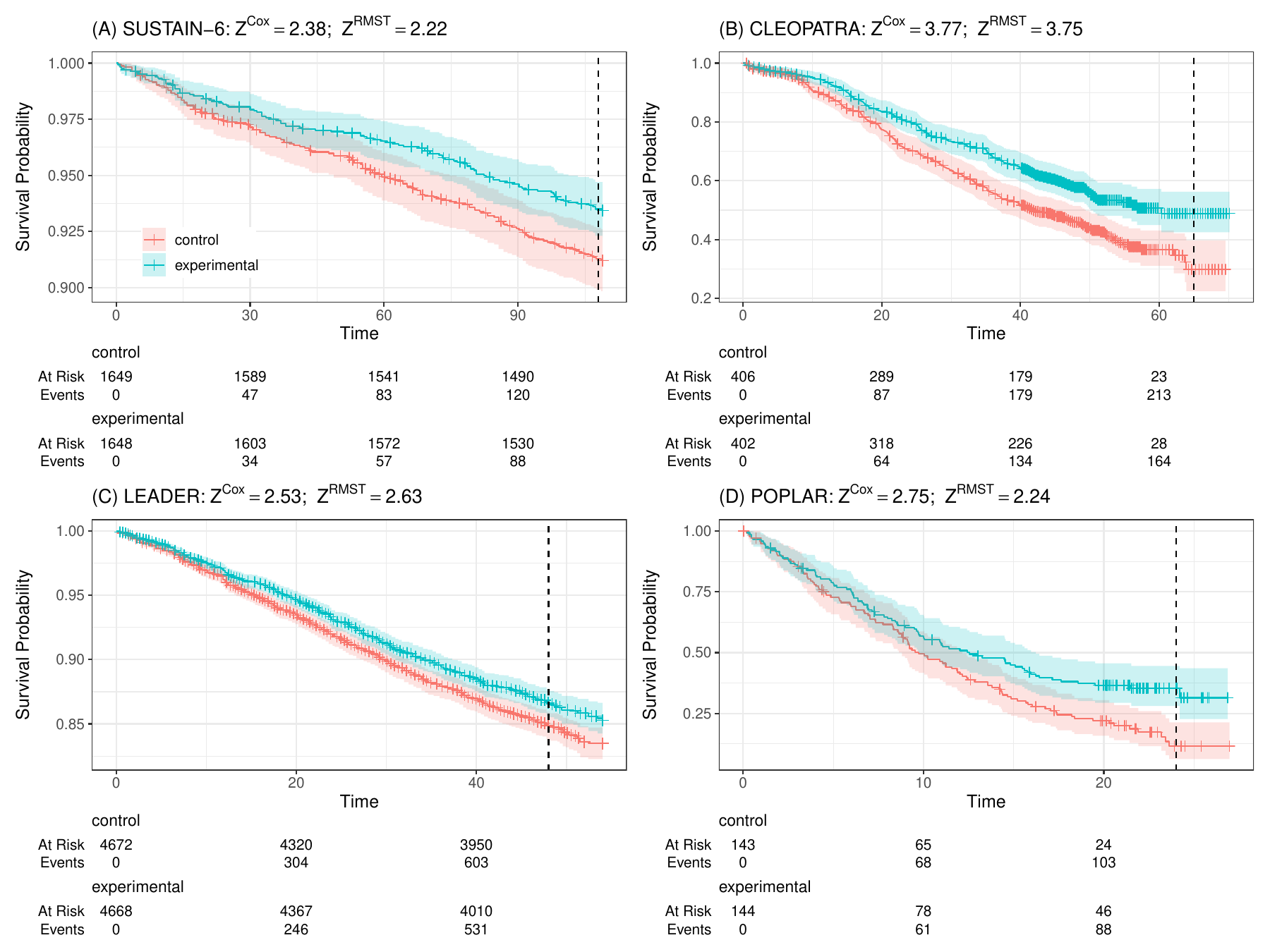}
\caption{\label{figKM} (Reconstructed-) Kaplan--Meier plots of example data sets. A: Low-event-rate-low-censoring trial with PH; B: High-event-rate-high-censoring trial with PH; C: Low-event-rate-low-censoring trial with early effect; and D: High-event-rate-low-censoring trial with a delayed effect. Dashed vertical lines correspond to the restriction time used in the case studies.} 
\end{figure*}

The first case study is based on the the SUSTAIN-6 trial \cite{marso2016semaglutide}, which compared the cardiovascular safety profile of semaglutide versus placebo in patients with type 2 diabetes. We recovered approximate patient-level data from the published Kaplan--Meier curve for the primary time-to-event endpoint using the algorithm of Guyot et al. \cite{guyot2012enhanced}. The approximate data set is summarized in Figure \ref{figKM}(A). The event rate is low, with estimated survival probabilities above 90\% at the end of the trial. Although recruitment was not necessarily fast, the trial was designed so that each patient was followed for 109 weeks, which produces a censoring pattern equivalent to an instant recruitment scenario in an event-driven trial.

The second case study is based on the CLEOPATRA trial \cite{swain2015pertuzumab}, which compared pertuzumab versus placebo as an add-on therapy in patients with HER2 positive metastatic breast cancer. Approximate patient-level overall survival data are available \textit{via} the \{kmData\} R package \cite{fell2021kmdata}, from which the Kaplan--Meier estimates are reproduced in Figure \ref{figKM}(B). The event rate is relatively high, with survival probabilities below 50\% at the end of follow-up. The censoring kicks in towards the end of follow-up but not extremely so, indicating a “modest” recruitment rate scenario.


The third case study is based on the LEADER trial \cite{marso2016liraglutide}, which like the SUSTAIN-6 trial was a large cardiovascular outcomes trial in patients with type 2 diabetes, this time comparing liraglutide versus placebo. We recovered approximate patient-level data from the published Kaplan--Meier curve for the primary time-to-event endpoint using the algorithm of Guyot et al. \cite{guyot2012enhanced}. The approximate data set is shown in Figure \ref{figKM}(C). The sample size is large and the event rate is low. The censoring kicks-in towards the end of follow-up, indicating a “fast” recruitment rate scenario.


The final case study is based on the POPLAR trial \cite{fehrenbacher2016atezolizumab}, which compared atezolizumab versus docetaxel for patients with non-small-cell lung cancer. Patient-level survival data are published in Gandara et al. \cite{gandara2018blood}, from which the Kaplan--Meier estimates are reproduced in Figure \ref{figKM}(D). The event rate is high. The trial duration is around 30 months and it is clear from the censoring pattern that recruitment took approximately 8 months: relatively fast.

The four case studies have been selected to provide a variety of event rates and recruitment rates. In each case, in order to estimate and visualize the hazard ratio function, we shall use a flexible parametric model with a spline on the cumulative hazards \cite{royston2001flexible}. Details are provided in the Supplementary Material. For each case study, we report the Z~statistics for the RMST-based test and the log-rank test. For the RMST-based tests we pick a value of $\tau$ close to the end of follow-up, as indicated in Figure \ref{figKM}. The idea is to characterize the trials according to their event rates, recruitment rates and also hazard ratio functions, and then observe whether or not the Z~statistics from each particular data set matches what we would expect based on our asymptotic and simulation results. 

We shall also use an additional visualization technique to deepen our understanding of the relative behaviour of the test statistics. Following Jiménez et al. \cite{jimenez2024visualizing}, both the log-rank statistic and a statistic asymptotically equivalent the RMST-based test statistic can be expressed as a difference in average “score” between the two arms, where each patient’s score is a mapping from their follow-up time and event status onto a one-dimensional scale. For a particular data set, plotting the scores for the two tests side-by-side provides additional insight into how much emphasis the tests place on early versus late follow-up periods.

\section{Results} \label{secResults}

\subsection{Weight functions}

\begin{figure*}
\centering
\includegraphics[width=\linewidth]{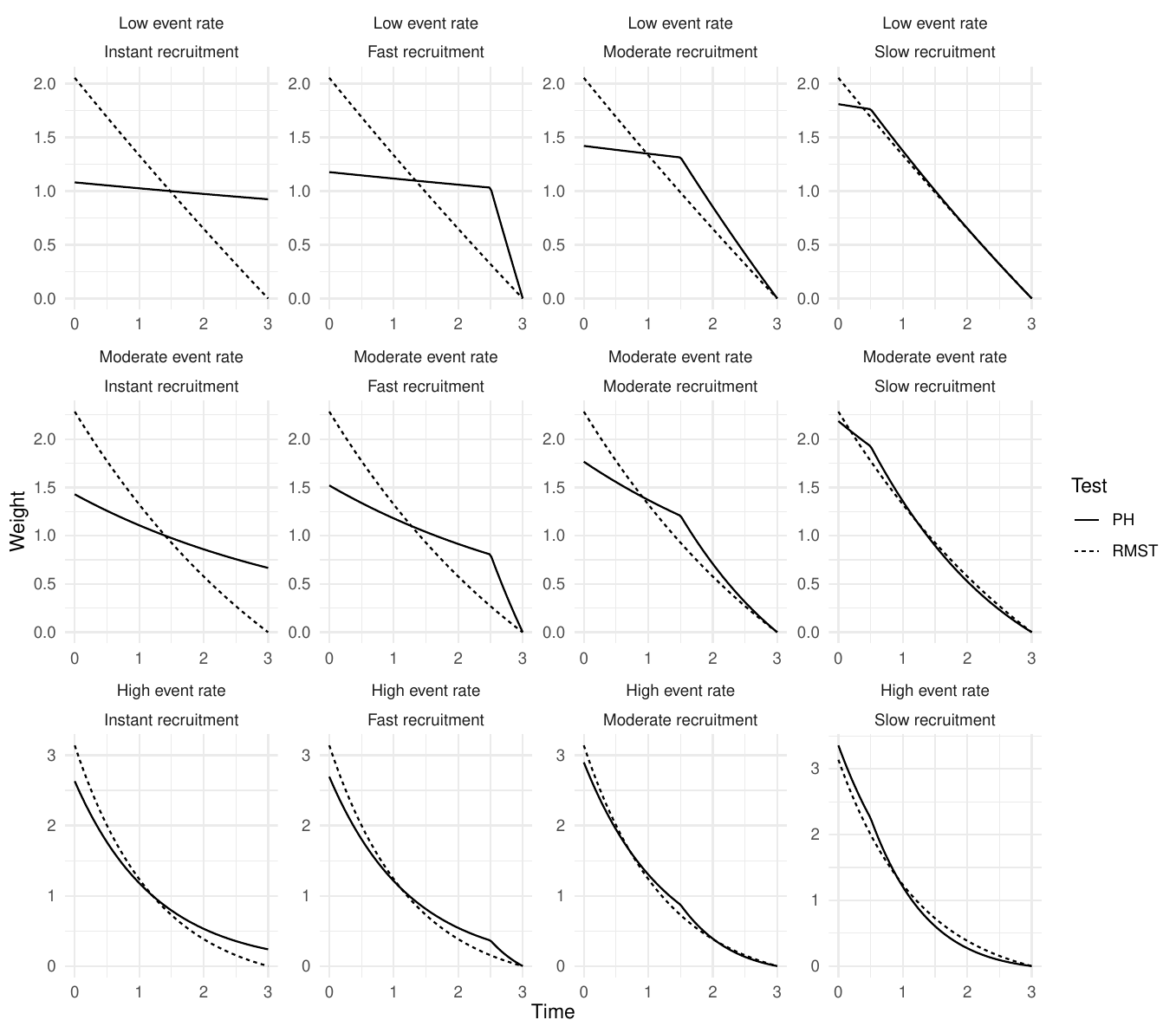}
\caption{\label{figWeights} Weight functions of RMST (\ref{eqRMST}), dashed lines, and PH-based (\ref{eqWD}), solid lines, tests, standardized to have mean equal to 1. The first row corresponds to scenarios 1 to 4 in Table \ref{tabScenarios}, the second row corresponds to scenarios 5 to 8, and the bottom row to scenarios 9 to 12. In almost all scenarios, the RMST-based test gives larger weight to earlier timepoints, compared with the log-rank (PH) test. However, it is only for scenarios with low event rates and fast recruitment rates that the difference between the two tests is noticeable. For scenarios with high event rates and/or slow recruitment, the two test have very similar weights when expressed in the form (\ref{eqRMST}) and (\ref{eqWD}). }
\end{figure*}

In Figure \ref{figWeights}, the weight functions (\ref{eqWD}) and (\ref{eqWtheta}) are plotted for all scenarios is Table \ref{tabScenarios}. The weights have been standardized to have mean 1 to ease the visual comparison. The RMST weights are unaffected by the censoring distribution, whereas the log-rank test gives relatively more weight to early time periods when censoring is heavy compared to when censoring is minimal. It is clear that when the event rate is high and/or recruitment is slow, the two tests are similar. This suggests relative efficiency should be close to one, not only under PH but also under other shapes of alternative hypothesis. When event rates are low and recruitment is fast, however, the RMST-based test is putting relatively more weight on the early part of follow-up compared to the PH analysis. Since the PH analysis is tailored for PH alternatives, we would expect that in these scenarios the higher early weighting of the RMST-based test should be less efficient.

\subsection{Simulation results}

Simulation results are shown in Table \ref{tabResults}. In general, for scenarios where both methods make use of the same amount of data, RMST has lower or equal power compared to the PH analysis. As expected from the asymptotic analysis based on the weight functions, differences in efficiency are small if there is a high event rate and/or there is a slow recruitment rate. For low/moderate event rates where the recruitment rate is fast, the efficiency of RMST can be considerably lower than PH analysis. In the scenarios we consider where the PH analysis makes use of more data, the efficiency gain can be large. We also report the mean (across simulations) percentage of events that occur after the restriction time, which varies across the scenarios but is generally very low. 

\begin{table*}[h]
\centering
\begin{tabular}{ | c c c |  c  c c c | c c  c  c |}
\hline
Scen- &                           &             & \multicolumn{2}{c}{Power} & Rel. Eff. & $\bar{\tau}$ & \multicolumn{2}{c}{Power} & Rel. Eff. + & \% events\\
ario & Event rate                & Recruitment & RMST  &  PH  &  & & RMST + &  PH +     &  &  $>\tau$ \\
\hline 
1 & \multirow{4}{*}{Low}      & Instant     & 0.79  & 0.88 & 0.77  &  3.00  & 0.79   & 0.92            & 0.66      &   13 \\
2 &                          & Fast        & 0.79  & 0.86 & 0.83  &  3.00  & 0.79   & 0.90            & 0.71      &    7\\
3 &                          & Moderate    & 0.77  & 0.79 & 0.94  &  3.00  & 0.78   & 0.86            & 0.82      &    3\\
4 &                          & Slow        & 0.69  & 0.69 & 1.00  &  3.00  & 0.76   & 0.79            & 0.92      &    2\\  
\hline
5 & \multirow{4}{*}{Moderate} & Instant     & 0.80  & 0.87 & 0.82   & 3.00  & 0.80   & 0.90            & 0.74      &    10 \\
6 &                          & Fast        & 0.80  & 0.85 & 0.88   & 2.99  & 0.80   & 0.89            & 0.77      &    5\\
7 &                          & Moderate    & 0.78  & 0.79 & 0.98   & 2.98  & 0.80   & 0.85            & 0.88      &     2\\
8 &                          & Slow        & 0.70  & 0.70 & 1.01   &  2.97 & 0.76   & 0.78            & 0.96      &     1\\
\hline
9 & \multirow{4}{*}{High}     & Instant     & 0.88  & 0.90 & 0.93   &  3.00 & 0.88   & 0.92            & 0.89      &    4 \\
10 &                          & Fast        & 0.88  & 0.89 & 0.95   &  2.96 & 0.88   & 0.91            & 0.91      &    2 \\
11 &                          & Moderate    & 0.86  & 0.87 & 0.99   & 2.89  & 0.88   & 0.89            & 0.96      &    1 \\
12 &                          & Slow        & 0.80  & 0.80 & 1.00   &  2.83 & 0.85   & 0.86            & 0.99      &    1 \\
\hline
\end{tabular}
\caption{\label{tabResults} Simulation results corresponding to the scenarios in Table \ref{tabScenarios}. RMST+, PH+ and Rel. Eff.+ refer to the fact that $t_H^+=3.5$ is the maximum follow-up time for the log-rank test instead of $t_H = 3$ ($\tau$ is still fixed at 3 years in these scenarios). Based on 10,000 repetitions, the Monte Carlo S.E. on estimated power of 0.9 is $\sqrt{\frac{0.9\times (1-0.9)}{10,000}}=0.003$.}
\end{table*}

\subsection{Case studies}

\begin{table*}
\centering
\begin{tabular}{ |l | c c c | c  c  | c c |}
\hline
Example & Recruitment & Event Rate & PH & Cox Z & RMST Z  & $\tau$ & \% events \\
 & & & & & & & after $\tau$ \\
 \hline
 A: SUSTAIN-6 & ``Instant'' &  Low  & Yes & 2.38   & 2.22 & 108 & 1 \\
 B: CLEOPATRA & Moderate    &  High & Yes & 3.77   & 3.75 & 65 & 0 \\
 C: LEADER & Fast        &  Low  & Early effect & 2.53 & 2.63 & 48 & 3 \\
 D: POPLAR & Fast        &  High & Late effect  &  2.75 & 2.24 & 24 & 1 \\
\hline
\end{tabular}
\caption{\label{tabExample} Applying the Cox PH and RMST-based tests to the example data sets from Figure \ref{figKM}}
\end{table*}

\begin{figure*}
\centering
\includegraphics[width=\linewidth]{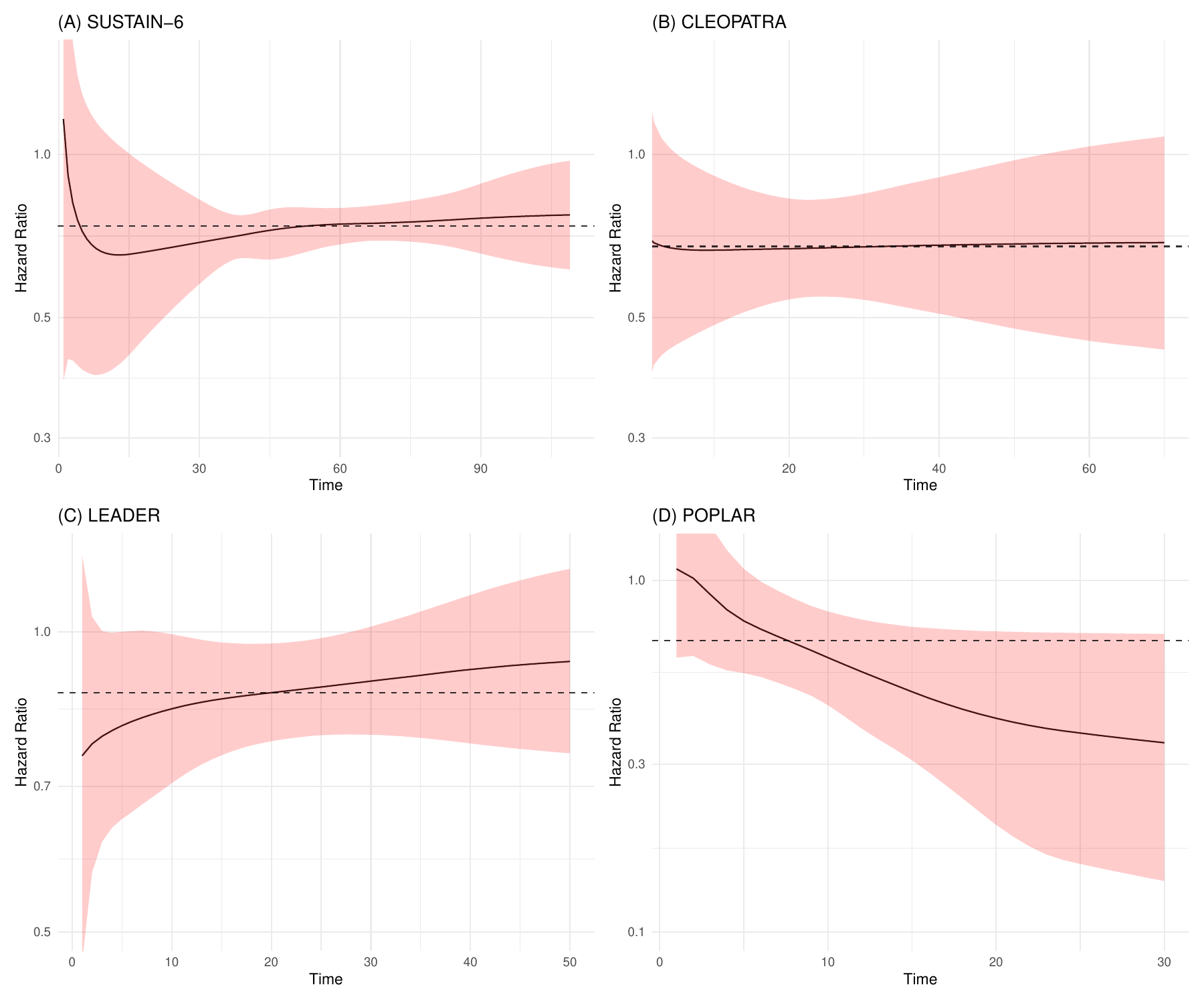}
\caption{\label{figHR} Estimated hazard ratios from example data sets. A: Low-event-rate-low-censoring trial with PH; B: High-event-rate-high-censoring trial with PH; C: Low-event-rate-low-censoring trial with early effect; and D: High-event-rate-low-censoring trial with a delayed effect. Dashed horizontal lines represent the estimated hazard ratio from a Cox model.}
\end{figure*}

For the first case study, based on the SUSTAIN-6 trial, the estimated time-varying hazard ratio appears to be relatively constant, as shown in Figure \ref{figHR}(A). Further visualizations of the model fit for all case studies are provided in Figures S1-S4 in the Supplementary Material. Categorizing this first case study as a low-event-rate-low-censoring trial with PH, then based on our asymptotic and simulation results we should expect the log-rank test to outperform the RMST analysis. This is indeed what we see from the Z~statistics in the first row of Table \ref{tabExample}.

\begin{figure*}
\centering
\includegraphics[width=\linewidth]{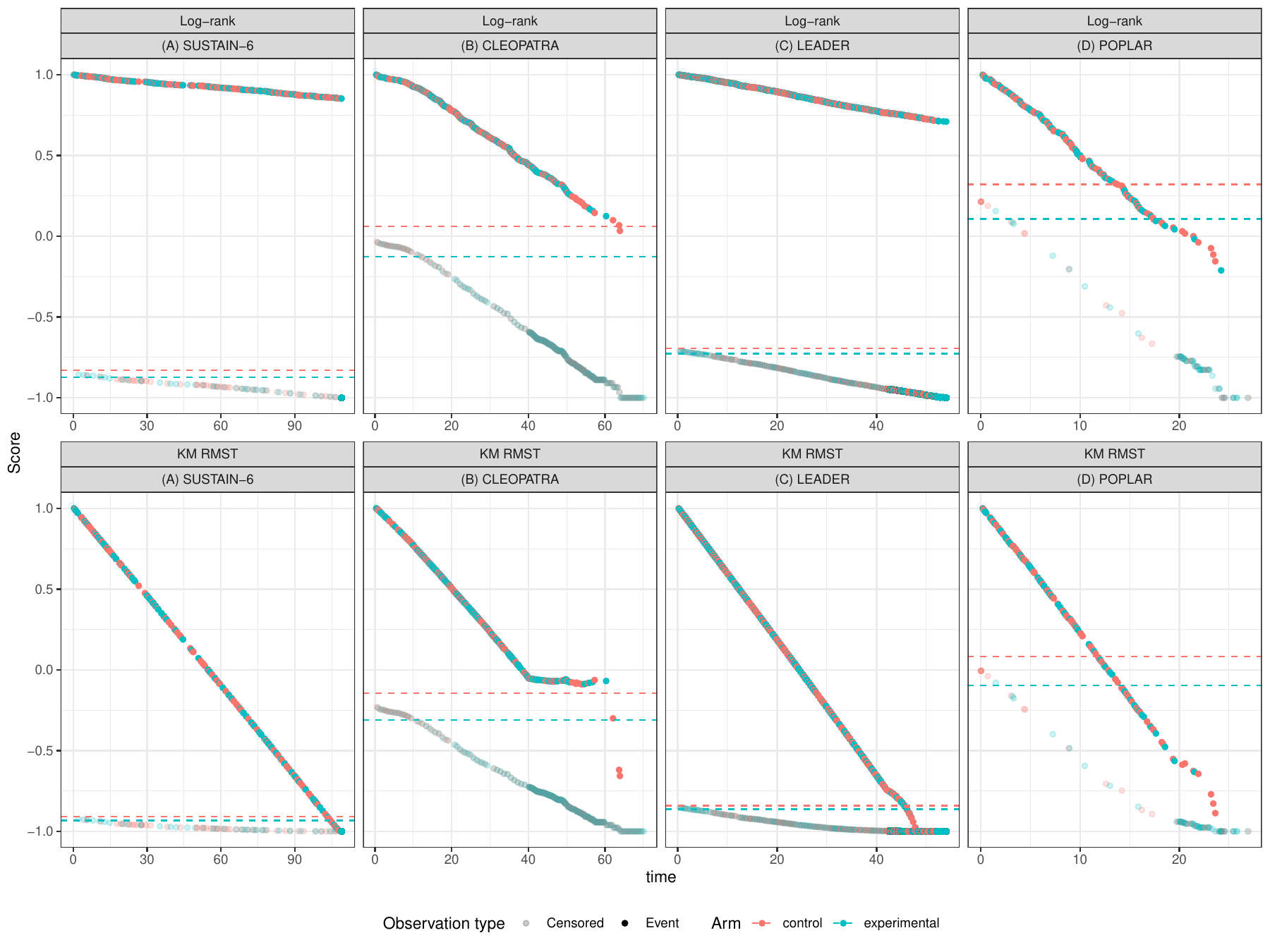}
\caption{\label{figScores} Standardized score plots of example data sets. Each dot corresponds to a patient in the trial, with x coordinate representing their follow-up time, y coordinate representing their derived score, and colour representing treatment group. Censored observations are partially transparent and are assigned lower scores than observed events given the same follow-up time. For all tests, patients with an early event are given the highest (worst) scores; patients who are censored after long follow up are given the lowest (best) scores. Horizontal dashed lines are plotted at the mean score on each arm. The distance between these two lines corresponds to the numerator in the permutation-based test statistic \cite{jimenez2024visualizing}. A: Low-event-rate-low-censoring trial with PH; B: High-event-rate-high-censoring trial with PH; C: Low-event-rate-low-censoring trial with early effect; and D: High-event-rate-low-censoring trial with a delayed effect}
\end{figure*}


For the second case study, based on the CLEOPATRA trial, the estimated time-varying hazard ratio appears remarkably flat, as shown in Figure \ref{figHR}(B). Categorizing this case study as a high-event-rate-moderate-censoring trial with PH, then based on our asymptotic and simulation results we should expect the RMST-based test and the log-rank test to perform similiarly. This is indeed what we see from the Z~statistics in the second row of Table \ref{tabExample}.


For the third case study, based on the LEADER trial, the estimated time-varying hazard ratio appears to be increasing slowly with time, as shown in Figure \ref{figHR}(C). This trial has a low event rate and fast recruitment rate, so under PH we would expect the log-rank test to outperform the RMST-based test. However, as can be seen from Figure \ref{figWeights}, the RMST-based test is giving greater weight to earlier timepoints compared to the log-rank test. If the hazard ratio is getting closer to one over time, then this works in favor of RMST-based test. We see from the third row of Table \ref{tabExample} that the RMST-based Z~statistic is larger in this case than the log-rank Z~statistic.


For the final case study, based on the POPLAR trial, the estimated time-varying hazard ratio is decreasing over time, as shown in Figure \ref{figHR}(D). This trial has a high event rate and fast recruitment rate, so under PH we would expect similar performance of the log-rank test and RMST-based test. We see, however, from the final row of Table \ref{tabExample} that the log-rank Z~statistic is much larger than the RMST-based Z~statistic. This demonstrates that small differences in relative weighting between early events and late events can have a big impact when the hazards are highly non-proportional.

Examining the “score” plots in Figure  \ref{figScores}, we see that the main qualitative difference between the log-rank and RMST-based tests is that, for any given follow-up time, the log-rank test always assigns a larger score to an event than to a censored observation, whereas for the RMST-based test, for follow-up times close to the restriction time, both events and censored observations are given a similar score. This is another way to understand that the RMST-based test is placing relatively less emphasis on the later follow-up compared with the log-rank test. The differences in the score plots (RMST \textit{vs}.\ log-rank) are less pronounced for the situations with high event rates and/or more censoring (B and D), compared to low-event-rate-high-censoring scenarios (A and C).

\section{Discussion}\label{secDiscussion}

Our review of existing asymptotic results, as well as our simulation results, have revealed that the relative efficiency of nonparametric RMST-based analysis compared to the log-rank test is indeed close to one in PH scenarios where the event rate is high and/or recruitment is slow, and when both methods use the same amount of data. This is consistent with several previous investigations \cite{tian2018efficiency, eaton2020designing, horiguchi2020empirical, royston2013restricted}. Such scenarios could occur in trials in aggressive and/or rare disease or in a competitive landscape where recruitment is challenging. However, when there is a low event rate and fast recruitment, the efficiency loss for RMST compared to PH-analysis is more considerable, even when both methods use the same amount of data. Such scenarios were considered by Freidlin et al. \cite{freidlin2021restricted}, and for this reason those authors reached different conclusions compared to much previous research. We might expect such scenarios to occur in large trials on indolent cancer or cardiovascular disease.

The choice of restriction time in the RMST-based analysis is a difficult issue. In particular, does it need to be explicitly pre-fixed? Royston \& Parmar \cite{royston2013restricted} make a proposal for a data-adaptive choice of restriction time which is based on minimizing the variance of the test statistic. A careful approach is needed, since it is clear that picking the restriction time in response to observed differences in survival could lead to severe inflation of error rates. Tian \textit{et al. }\cite{tian2020empirical} have shown that, from a technical perspective, it is sometimes possible to make valid inference on RMST even when the restriction time is equal to the last observation time (the validity of the large sample approximation of the test statistics' sampling distribution depends on aspects of the recruitment process that might be difficult to verify ahead of time). In many empirical comparisons of existing data sets, $\tau$ is chosen retrospectively at a value as close as possible to $t_H$ \cite{horiguchi2020empirical}. From a practical perspective, however, the restriction time is more likely to be pre-fixed to an easily digestible round number, such as 12 months or 48 months. This means that the final data set will often contain events collected after the restriction time that could potentially be used in a PH analysis, but not in a nonparametric analysis of the RMST based on the pre-fixed restriction time. Taking this aspect into consideration, the efficiency loss of moving to the RMST-based analysis becomes greater, but it is difficult to quantify how much greater.

We restricted our simulation scenarios to proportional hazards scenarios, as this was the area of conflict from previous investigations, requiring clarification. There is of course no fundamental reason to expect proportional hazards in general. Our case studies (C) and (D) illustrated how non-PH in the data further impact relative efficiency. 

The key takeaway from our research is not that the log-rank test should be preferred to the RMST-based test when we anticipate PH. There are good arguments for moving towards model-free estimands and assumption-lean analysis methods in RCTs. Rather, the key takeaway is that such a movement may come with an efficiency cost under PH scenarios, contrary to what has previously been suggested.

Apart from RMST, other model-free summary measures, such as milestone probabilities, could be used to construct model-free assumption-lean analysis approaches. Tests that are based on nonparametric estimation of milestone probabilities may have competitive power in some scenarios \cite{chen2023beyond}, but generally they are vulnerable to power loss because they involve dichotomization of a continuous variable \cite{senn2009measurement}. More research is needed to characterize those situations where power loss is likely to be low.

Another key area of future research is to evaluate the relative efficiency of assumption-lean analysis approaches that make use of covariate adjustment, compared to proportional hazards methods. This is an active area of research \cite{ozenne2020estimation, diaz2019improved, chen2023beyond}. It is important to thoroughly understand the advantages and disadvantages of such proposals prior to implementation. Such an evaluation is doomed from the start if we do not fully understand their relative efficiency in a simpler situation without covariates.

\section{Acknowledgements}
TPM is funded by the UK Medical Research Council (grant MC\_UU\_00004/07).

\bibliographystyle{ieeetr}
\bibliography{references}

\end{document}